# Layer- and Field-Dependent Magnetic Order in 2D CrSBr Revealed by Pulsed Nanocalorimetry

Hugo Gomez-Torres[1,2], Roop K. Mech[3], Alessandra Canetta[3], Llibertat Abad[4], Carles Navau[2], Daniel G. Chica[5], Xavier Roy[5], Pascal Gehring[3,6], Kenji Watanabe[7], Takashi Taniguchi[8], Aitor Lopeandia[1,2*], Javier Rodriguez-Viejo[1,2*]

[1] Catalan Institute of Nanoscience and Nanotechnology (ICN2), CSIC and BIST, E-08193, Bellaterra, Spain.

[2] Departament de Física, Universitat Autònoma de Barcelona, E-08193, Bellaterra, Spain.

[3] Institute of Condensed Matter and Nanosciences, Université Catholique de Louvain (UCLouvain), 1348 Louvain-la-Neuve, Belgium.

[4] Institut de Microelectrònica de Barcelona (IMB-CNM-CSIC). Campus de la UAB, E-08193, Cerdanyola del Vallès (Barcelona), Spain.

[5] Department of Chemistry, Columbia University, New York, New York 10027, United States.

[6] WEL Research Institute, Avenue Pasteur 6, 1300 Wavre, Belgium.

[7] Research Center for Electronic and Optical Materials, National Institute for Materials Science, 1-1 Namiki, Tsukuba 305-0044, Japan

[8] Research Center for Materials Nanoarchitectonics, National Institute for Materials Science, 1-1 Namiki, Tsukuba 305-0044, Japan

*Correspondence: Javier Rodriguez-Viejo (javier.rodriguez@uab.cat), Aitor Lopeandia (aitor.lopeandia@uab.cat)

Understanding the evolution of magnetic order in the two-dimensional limit remains a central challenge in van der Waals magnets, where thermodynamic measurements are constrained by the femtogram-scale mass of exfoliated flakes. Here, microsecond pulse-heating nanocalorimetry is used to measure the heat capacity and magnetic entropy of CrSBr flakes down to the monolayer limit. The measurements reveal the entropy landscape associated with magnetic ordering, uncovering a decreasing interlayer transition temperature toward the monolayer limit and an entropy-derived effective moment per layer that increases with thickness toward the expected spin-only scale. Thermodynamic anomalies capture a crossover from bulk-like interlayer antiferromagnetism to a regime dominated by intralayer ferromagnetic correlations. A pronounced layer-parity effect further emerges, with odd-layer samples displaying an additional high-temperature contribution associated with uncompensated magnetic layers. Under in-plane magnetic fields applied along the easy axis, antiferromagnetic order is progressively suppressed, allowing extraction of a thickness-dependent characteristic suppression field reflecting weakened interlayer exchange coupling. Entropy analysis further reveals an extended regime of magnetic fluctuations persisting well above the interlayer ordering transition. Together, these results establish nanocalorimetry as a powerful thermodynamic probe of low-dimensional magnetism, providing direct access to magnetic entropy, exchange interactions, and dimensional crossover in atomically thin van der Waals magnets.

## Introduction

The emergence of intrinsic magnetism in atomically thin van der Waals (vdW) crystals has enabled the exploration of low-dimensional spin phenomena and expanded the prospects for ultrathin spintronic devices.[1–7] Compared with conventional three-dimensional magnets, van der Waals magnets are quasi-two-dimensional systems in which weak interlayer exchange and magnetic anisotropy stabilize the ordered state and render the magnetic ground state highly tunable by external stimuli such as gating, strain or magnetic field.[8–11] A central question in the field is how the thermodynamic stability of these ordered states evolves as the thickness is reduced toward the two-dimensional limit.

Among vdW magnets, chromium sulfide bromide (CrSBr) is a model platform for studying low-dimensional magnetism in semiconducting systems.[12–18] CrSBr combines ambient robustness and semiconducting behavior with pronounced in-plane magnetic anisotropy arising from its mixed-anion coordination environment.[12–14,18] In contrast to highly air-sensitive halide magnets such as $CrI_3$, and to metallic vdW ferromagnets such as $Fe_3GeTe_2$, CrSBr provides a chemically robust semiconducting platform for studying anisotropic magnetism without severe degradation or metallic screening. An additional practical advantage is that mechanically exfoliated CrSBr flakes cleave along crystallographic axes, allowing the magnetic easy axis to be directly identified from the flake geometry. Its magnetic order is strongly coupled to excitonic and optical properties, enabling magnetic control of excitonic transitions, exciton-coupled coherent magnons, and magnetically tunable exciton–polariton responses.[19–21] CrSBr is an A-type antiferromagnet: within each layer, spins align ferromagnetically along the easy b-axis, while adjacent layers couple antiferromagnetically through weak interlayer exchange, yielding a bulk Néel transition near 132 K. In addition to this long-range antiferromagnetic transition, CrSBr hosts pronounced magnetic anisotropy, dynamic magnetic crossovers, and persistent intralayer ferromagnetic correlations that survive to higher temperatures around 145-160 K.[22–25] These features position CrSBr as an ideal system for probing how magnetic order, correlations, and entropy evolve as the material is thinned toward the two-dimensional limit.

A variety of techniques, including magnetotransport,[23,26,27] magneto-optical probes,[12,28] and local magnetic imaging methods such as MFM,[29] NV magnetometry[30–32] and scanning SQUID[11,33] have established the quasi-two-dimensional nature of CrSBr, revealed spin-reorientation processes, and quantified its magnetic anisotropy and field-driven phase behavior. Recent thermoelectric measurements have further shown that spin entropy can strongly affect the transport response of CrSBr,[34] highlighting the relevance of entropy-related magnetic degrees of freedom in this material. Heat-capacity measurements have previously been reported for bulk CrSBr,[12,28,35] and a nanomechanical approach has demonstrated the detection of a heat-capacity anomaly in a microscale CrSBr membrane.[36] However, no systematic, layer-resolved calorimetric investigation of CrSBr has yet been reported across the few-layer regime as a function of thickness and applied magnetic field. Because exfoliated flakes contain only femtograms of material, well below the sensitivity of conventional calorimetry, accessing this layer- and field-dependent thermodynamic evolution remains challenging.

Here we overcome this limitation by employing microsecond Pulse-Heating nanocalorimetry (µs-PHnC)[37,38] to measure the heat-capacity of individual CrSBr flakes down to the monolayer. With sub-pJ/K sensitivity and microsecond time resolution, µs-PHnC allows us to track the specific heat as a function of temperature, magnetic field and layer number. We resolve the thermodynamic anomaly associated with intralayer ferromagnetism and interlayer antiferromagnetic order in exfoliated CrSBr, and quantify its shift and suppression under an in-plane magnetic field. This reveals a dramatic decrease in the corresponding characteristic suppression field as the material approaches the 2D limit.

Beyond locating the transition, µs-PHnC provides direct access to the magnetic entropy by integrating the excess heat capacity $\Delta C_p/T$. This offers a quantitative thermodynamic complement to magnetotransport, Raman spectroscopy, magneto-optical probes and local magnetic imaging, which primarily track transport, excitations, symmetry breaking or local magnetic fields rather than the entropy released across the transition. The measurement therefore links the observed magnetic phase behavior to the entropy released at ordering, enabling quantitative comparison of magnetic correlations across thicknesses and providing a thermodynamic route to infer the effective magnetization per layer or per unit area.

## Results

CrSBr crystallizes in a FeOCl-type orthorhombic structure with space group Pmnm, consisting of van der Waals layers of edge-sharing [$CrS_4Br_2$] octahedra stacked along the crystallographic c-axis, as shown in Figure 1a. Within each layer, $Cr^{3+}$ ions form a quasi-square magnetic lattice. As introduced above, the magnetic ground state is A-type antiferromagnetic, with ferromagnetically ordered layers aligned along the easy b-axis and weak antiferromagnetic coupling between adjacent layers.

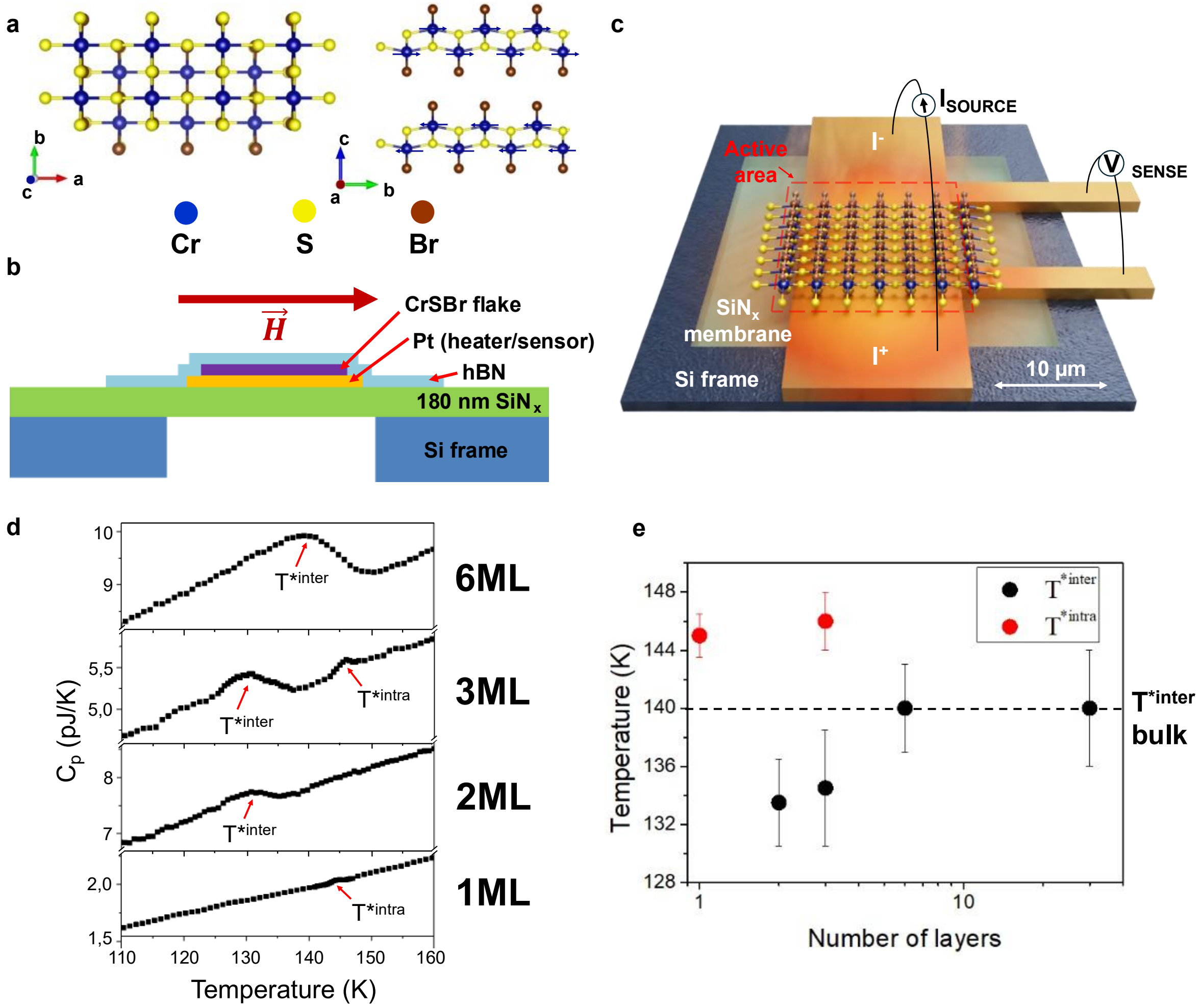

**Figure 1. Structure, device, and layer-dependent heat capacity of CrSBr.** (a) Crystal structure of CrSBr along [001] direction (left panel) and [100] (right panel). b) Schematic of the suspended Pt/SiN nanocalorimetric device. (c) Illustration of the experimental configuration for microsecond pulsed-heating nanocalorimetry. (d) Heat capacity $C_p$ (T) of CrSBr flakes with different layer numbers. (e) $T^{*inter}$ and $T^{*intra}$ temperatures extracted from the inflection points of the $C_p$ (T) curves as a function of layer number.

Figures 1b-c show a schematic cross-section of the nanocalorimetric chip and the type of device employed in this work, respectively. The device consists of a suspended $SiN_x$

membrane (180 nm thick) patterned with a Pt strip that simultaneously acts as heater and thermometer in a four-probe configuration. Exfoliated CrSBr flakes, encapsulated with hBN layers to prevent degradation[34], are transferred onto the Pt heater/sensor. The use of hBN encapsulation preserves the intrinsic magnetic and electronic properties of exfoliated CrSBr. Its high in-plane thermal conductivity also enhances thermal coupling between the heater and the sample. The device is supported by a Si frame that provides mechanical stability while the suspended membrane geometry reduces the thermal conductance to the substrate and minimizes parasitic heat leakage during the microsecond heating pulse. An in-plane magnetic field (*H*) is applied along the device, aligned with the easy b-axis of CrSBr. The devices are fabricated using PVC polymer dry-transfer technique[39] inside a $N_2$-filled glovebox, ensuring precise positioning and intimate thermal contact with the Pt heater/thermometer. A detailed description of the device fabrication, measurement protocol, and experimental conditions is provided elsewhere,[37] and the key experimental details are summarized in the Methods section.

Heat-capacity measurements were performed using microsecond pulsed-heating nanocalorimetry under high vacuum (~$10^{-8}$ mbar) with a base temperature stability of ~1 mK. A current pulse of 50 µs duration, generated by a Howland current source (12–17 mA depending on device size), was applied to the Pt heater, producing a heating rate of ~$10^5$ K/s, with a time-off of 0.1 s between pulses to ensure full thermal relaxation. The resulting voltage signal was acquired in a differential configuration to maximize sensitivity. The temperature increase was extracted from the calibrated resistance $R(T)$ of the Pt sensor. Heat capacity was determined from the linear regime of the temperature transient under nearly adiabatic conditions. A full derivation of the heat-capacity expression, including the differential reference subtraction, is described in Ref.[37]

Figure 1d presents the temperature-dependent heat capacity, $C_p$ (*T*), of CrSBr flakes with varying thickness. After subtracting the addenda, the curves were further adjusted to account for device-to-device variations coming from the hBN encapsulation. This relative baseline correction was estimated from the reported heat capacity of hBN at the measurement temperature[40–42] and from the measured hBN dimensions of each device. Details of the offset analysis are provided in the Supplementary Information.

Despite the overall phonon-dominated increase of $C_p$ with temperature, well-defined anomalies are observed in the vicinity of the magnetic ordering temperatures. The low-*T* anomaly points to the onset of the interlayer antiferromagnetic ($T^{*inter}$) transition in CrSBr. The persistence of $C_p$ anomalies down to the monolayer indicates that magnetic ordering in CrSBr remains detectable even in the two-dimensional limit.

The transition observed in the 6 ML and bulk-like flakes appears around 140 K, slightly above the $T_N$ ~132 K commonly reported by conventional bulk calorimetry. Similar transition temperatures near 139 K have also been reported in previous heat-capacity studies of bulk-like CrSBr flakes.[36] This difference may arise from strain induced during device transfer, as well as from the high heating rate involved in the microsecond pulse-heating protocol (~$10^5$ K/s), which may shift the apparent position of the calorimetric anomaly. The conclusions of this work rely on the relative evolution of the thermodynamic anomaly with layer number and magnetic field, measured under the same protocol.

In the monolayer limit, $C_p(T)$ evolves smoothly up to a weak anomaly at ∼145 K, indicating the persistence of intralayer ferromagnetic correlations ($T^{*intra}$) even in the absence of long-range interlayer order.

The transition temperatures extracted from the inflection points of $C_p(T)$ are summarized in Figure 1e, providing a model-independent measure of the thickness evolution of magnetic order. $T^{*inter}$ decreases monotonically with decreasing layer number, consistent with weakened interlayer antiferromagnetic exchange under dimensional confinement and with theoretical and experimental descriptions of finite-size effects in A-type antiferromagnets.[43] Similar reductions of $T^{*inter}$ together with a broadening of the magnetic heat-capacity anomaly, have been observed in other ultrathin magnetic systems.[44,45] This thermodynamic trend with thickness differs from recent second harmonic generation (SHG) studies of few-layer CrSBr,[12,28] which reported higher magnetic onset temperatures when decreasing thickness. The apparent discrepancy likely reflects the different quantities probed by the two techniques. SHG is sensitive to symmetry breaking and can detect local or surface magnetic order; indeed, recent SHG work identified a surface-related magnetic transition near 140 K in the monolayer limit, distinct from the bulk antiferromagnetic transition around 132 K. In contrast, heat capacity measures the temperature range over which magnetic entropy is released. Our nanocalorimetry data therefore indicate that the main thermodynamic anomaly associated with long-range interlayer antiferromagnetic order shifts to lower temperatures as the layer number is reduced, even though short-range, local, or surface-enhanced magnetic correlations may persist to higher temperatures and remain visible to optical probes. These results support a picture in which CrSBr hosts multiple magnetic temperature scales. In the few-ML limit, a higher-temperature magnetic correlation or symmetry-breaking regime is detected by symmetry-sensitive optical techniques, whereas a lower-temperature thermodynamic anomaly is resolved by calorimetry.

Beyond this dimensional crossover, µs-PHnC reveals a pronounced layer-parity effect. Odd-layer flakes show a clear additional heat-capacity anomaly, near T ≈ 145 K, whereas even-layer samples exhibit only a much weaker and broader contribution in the same temperature range. We assign this feature to intralayer ferromagnetic correlations, which develop within individual CrSBr layers independently of long-range interlayer antiferromagnetic order. Because CrSBr consists of ferromagnetic layers coupled antiferromagnetically along the out-of-plane direction, even-layer stacks are magnetically compensated, whereas odd-layer stacks retain an uncompensated magnetic moment. Thus, although intralayer correlations are expected to emerge across all thicknesses, their calorimetric visibility is layer-parity dependent: in even-layer flakes they contribute mainly as a broad background variation in $C_p(T)$, whereas in odd-layer flakes the uncompensated moment enhances their thermodynamic signature, giving rise to a sharper high-temperature anomaly. This interpretation is consistent with magneto-optical and local-probe studies of CrSBr, which report finite net magnetization in odd-layer samples and compensated A-type antiferromagnetic stacking in even-layer samples.[27,29,31,32,46] The observation of this parity-dependent calorimetric response provides direct thermodynamic evidence of layer-resolved magnetic stacking and its associated uncompensated magnetic states in CrSBr.

**Magnetic-field dependence and suppression of interlayer order**

Figure 2 presents the magnetic-field dependence of the $C_p(T)$ of CrSBr flakes with different thicknesses, measured with an in-plane magnetic field applied along the crystallographic easy $b$-axis. Panels (a–d) correspond to 3ML, 2ML, 1ML, and 6ML devices, respectively, providing direct thermodynamic insight into the evolution of magnetic order from the bulk-like regime to the two-dimensional limit.

We begin with the 3ML sample, Figure 2a, which exhibits the most complex behavior due to its odd-layer stacking. In zero field, the $C_p(T)$ curve shows a broad anomaly centered around $T$~134 K, associated with the onset of interlayer antiferromagnetic correlations, together with a subtle high-temperature shoulder near 145 K attributed to intralayer ferromagnetic correlations. When a magnetic field is applied along the easy $b$-axis, the antiferromagnetic contribution is progressively suppressed, whereas the high-temperature intralayer-related feature remains essentially unchanged throughout the explored field range. This contrasting field response indicates that the applied field is sufficient to perturb the weak interlayer antiferromagnetic order, while remaining negligible with respect to the energy scale of the intralayer ferromagnetic exchange interactions. Consequently, the heat-capacity profile evolves in a characteristic manner, reflecting the distinct field sensitivities of the interlayer and intralayer magnetic interactions.

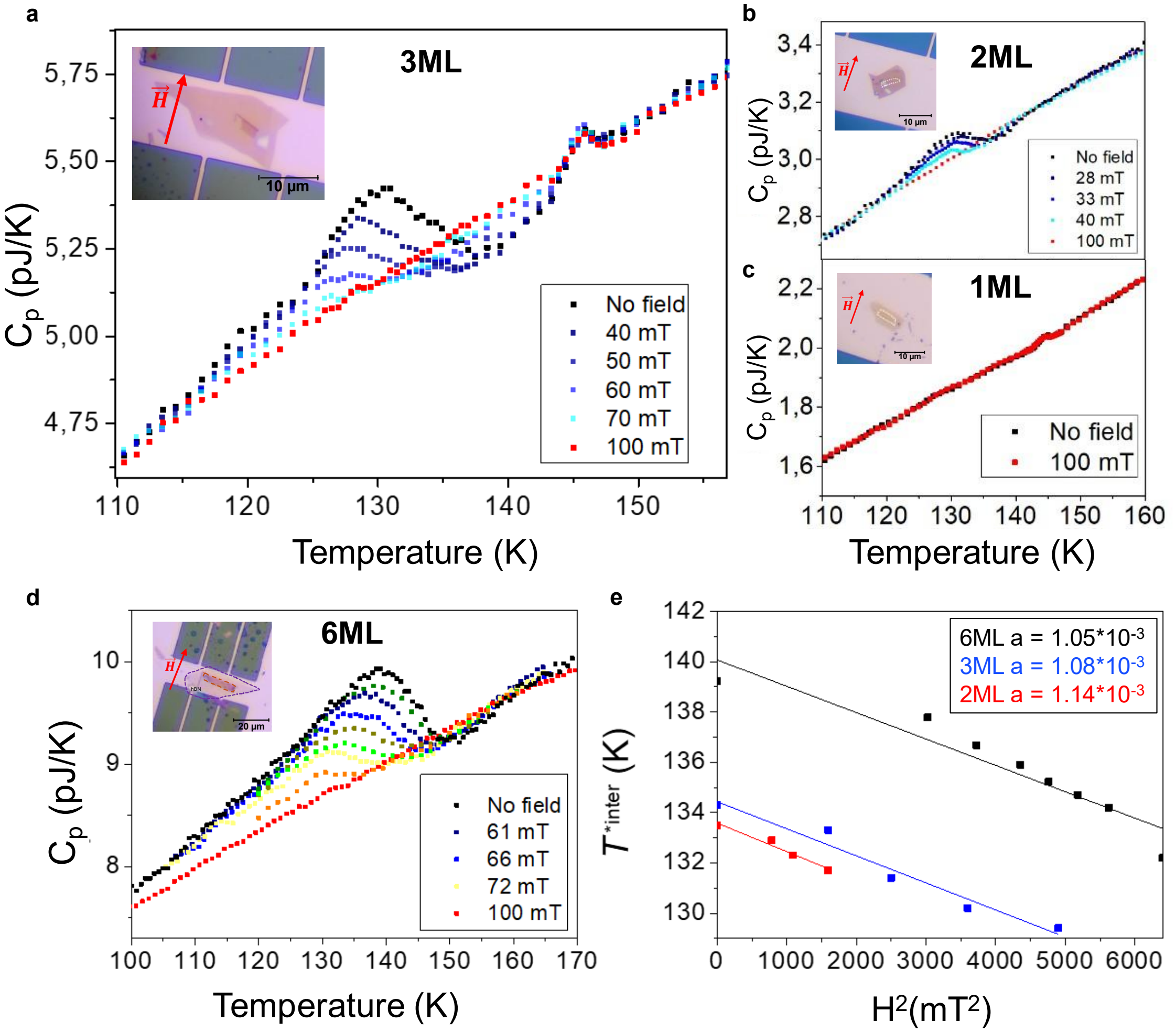

**Figure 2. Magnetic-field dependence of the heat capacity in few-layer CrSBr.** (a-d) Temperature-dependent heat capacity $C_p(T)$ of few-layer CrSBr of different thicknesses under in-plane magnetic fields applied along b-axis. (e) Extracted $T^{*inter}$ as a function of $H^2$ displaying a linear dependence consistent with $T^{*inter}(H) = T^{*inter}(0) - aH^2$.

The 2ML sample, Figure 2b, provides a direct comparison to the odd-layer response. As an even-layer stack, it forms a fully compensated antiferromagnetic state with no net magnetization. Under applied field, the anomaly is progressively suppressed. No clearly resolved high-temperature anomaly can be distinguished above the AFM transition. This does not eliminate the presence of intralayer ferromagnetic correlations in even-layer stacks, rather, their calorimetric contribution is expected to be broad and weak,[28,35] making them difficult to resolve over the phonon background and the experimental noise. In contrast, in odd-layer flakes, the uncompensated moment enhances the visibility of this contribution, producing a sharper high-temperature anomaly. The parity-dependent response is best understood as a difference in the calorimetric visibility of intralayer correlations, rather than as direct evidence for their absence in even-layer samples. In the monolayer limit (Figure 2c), where interlayer antiferromagnetic order is absent, the remaining intralayer-related heat-capacity feature at T ≈ 145 K shows no measurable shift or change under magnetic fields up to 100 mT.

The field dependence of the 6ML sample (Figure 2d) closely resembles the behavior observed in thinner multilayers. As the in-plane magnetic field increases, the antiferromagnetic anomaly progressively shifts to lower temperature, broadens, and decreases in amplitude. At the highest fields, the anomaly evolves into a smooth background, consistent with the progressive suppression of long-range interlayer antiferromagnetic order.

Overall, the field response of the multilayer samples reflects the different energy scales of interlayer antiferromagnetic and intralayer ferromagnetic interactions. The applied field primarily destabilizes the weak interlayer antiferromagnetic configuration, reducing the calorimetric anomaly and shifting the interlayer transition to lower temperature, while the intralayer-related contribution remains comparatively insensitive within the accessible field range.

To quantify this evolution, Figure 2e shows the $T^{*inter}$, extracted from the inflection point of $C_p(T)$, as a function of $H^2$ for all samples exhibiting interlayer antiferromagnetic order. In all cases, $\mathrm{T}^{*\mathrm{inter}}$ decreases linearly with $H^2$, consistent with the Landau-type quadratic suppression $\mathrm{T}^{*\mathrm{inter}}(H) = \mathrm{T}^{*\mathrm{inter}}(0) - aH^2$ expected for a collinear antiferromagnet.[47] The extracted coefficients are $a = (1.05 \pm 0.14) \times 10^{-3}$, $(1.08 \pm 0.13) \times 10^{-3}$, and $(1.14 \pm 0.14) \times 10^{-3}$ K·mT$^{-2}$ for 6ML, 3ML, and 2ML, respectively, with high-quality linear fits ($R^2 \sim 0.95$-$0.9$). Although the nominal values of *a* show a slight increase with decreasing thickness, this variation is smaller than the corresponding experimental uncertainty. Therefore, the extracted coefficients should be regarded as comparable within error, and no systematic thickness dependence of *a* can be concluded from these data.

Figure 3 illustrates the temperature dependence of the magnetic entropy change $\Delta S(T)$, for CrSBr flakes with different thicknesses and under applied magnetic field. The magnetic contribution is isolated by subtracting a Debye-type phonon background from the measured heat capacity, yielding the excess heat capacity $\Delta C_p(T)$ (see Methods and Supplementary Information). The magnetic entropy change is then calculated as $\Delta S(T) = \int (\Delta C_p/T)\, dT$, providing direct thermodynamic access to the magnetic degrees of freedom involved in the transition. In Figure 3a, $\Delta S(T)$ is normalized by flake area and number of layers, enabling a direct comparison of the thermodynamic contribution per layer. Although the absolute entropy values, and the entropy-derived effective moments, depend slightly on the chosen phonon background and hBN baseline correction, the relative trends with layer number and magnetic field remain robust under reasonable variations of the analysis parameters. Because $\Delta S(T)$ is an accumulated quantity, field-dependent entropy values are compared over equivalent temperature windows or at common reference temperatures to avoid apparent differences arising from unequal integration ranges.

The magnetic entropy evolves systematically with thickness, but not as a simple monotonic reduction of the normalized entropy. Instead, reducing the layer number primarily redistributes the entropy release in temperature. In the more bulk-like 6ML flake, $\Delta S(T)$ accumulates gradually over a broad temperature interval, consistent with magnetic correlations extending over a wide range around the antiferromagnetic transition. In 3ML, the entropy release becomes more concentrated near the reduced $T^{*inter}$, indicating that a larger fraction of the recovered magnetic entropy is released within a narrower temperature window. In contrast, the 2ML sample exhibits a broader and more gradual entropy evolution, suggesting that thermal fluctuations play an increasingly important role as the interlayer antiferromagnetic coupling is reduced in the bilayer limit. Together with the monotonic reduction of $T^{*inter}$ discussed in Figure 1e, these trends point to a weakening of interlayer antiferromagnetic coupling under dimensional confinement.

In all samples, a significant fraction of the magnetic entropy is released over a broad temperature range extending across and above the $T^{*inter}$, indicating that magnetic correlations develop progressively rather than emerging abruptly at the transition. This indicates that $T^{*inter}$ should not be interpreted as the first appearance of local magnetic correlations, but rather the establishment of interlayer coherence within a pre-correlated magnetic state.

The field dependence of $\Delta S(T)$ further highlights this redistribution. In the 6ML flake (Figure 3b) the entropy decreases with increasing field, consistent with the destabilization of interlayer antiferromagnetic correlations. In 3ML and 2ML (Figures 3c,d) the field response reflects the same balance between weakened interlayer coupling and layer parity discussed above, leading to a redistribution rather than a uniform quenching of the magnetic entropy.

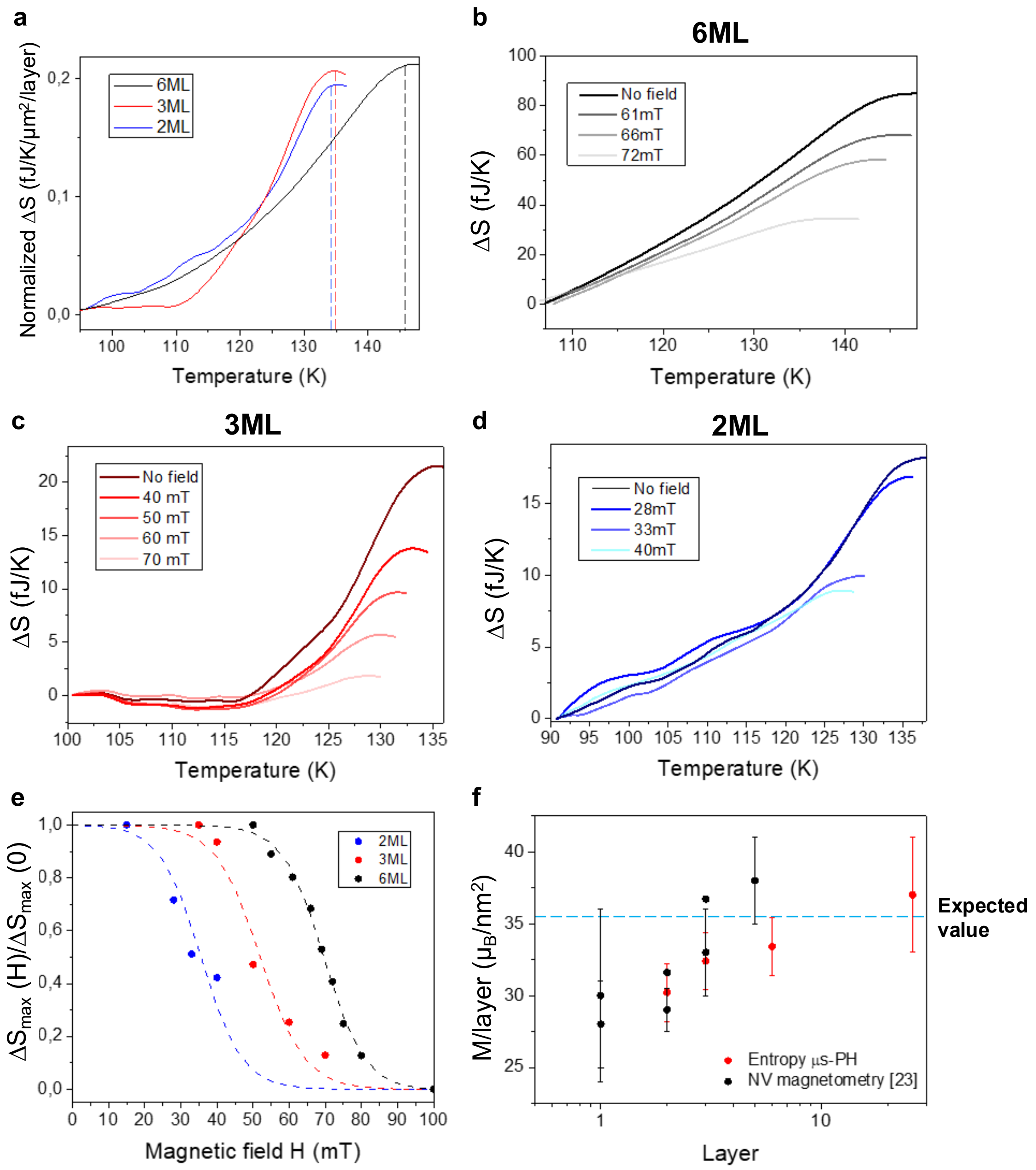

**Figure 3. Magnetic entropy and field-driven suppression of antiferromagnetic order in few-layer CrSBr.** (a) Temperature dependence of the magnetic entropy change $\Delta S(T) = \int (\Delta C_p/T)\, dT$, normalized by flake area and number of layers, for CrSBr flakes of different thicknesses. (b–d) Field dependence of ΔS(T) for (b) 6 ML, (c) 3 ML, (d) 2 ML samples under in-plane magnetic field along the crystallographic easy $b$-axis. (e) Normalized entropy suppression, $\Delta S_{max}(H)/\Delta S_{max}(0)$, as a function of magnetic field for 2 ML, 3 ML, and 6 ML flakes, highlighting the layer-dependent reduction of magnetic entropy. (f) Entropy-derived effective magnetic moment per layer estimated from the recovered magnetic entropy. Blue dashed line shows the expected bulk value.

The magnetic entropy complements the heat-capacity analysis by providing an integrated and fluctuation-sensitive probe of magnetic order. While clear anomalies in $C_p(T)$ identify the transition temperatures, $\Delta S(T)$ captures the strength and character of the transition, enabling a direct comparison of magnetic correlations across thickness and applied magnetic field. This links the observed shift and suppression of the calorimetric anomalies to a progressive weakening of interlayer antiferromagnetic coherence in the few-layer limit.

Figure 3e shows the normalized maximum magnetic entropy $\Delta S_{max}(H)/\Delta S_{max}(0)$ as a function of in-plane magnetic field for different thicknesses. For each field, $\Delta S$ is evaluated at the maximum accumulated entropy within the same integration window and normalized to the corresponding zero-field value. For all antiferromagnetic flakes, $\Delta S_{max}(H)$ decreases with field, reflecting the progressive suppression of the entropy associated with interlayer antiferromagnetic order. The suppression occurs over a well-defined field range and was quantified by fitting the normalized entropy suppression, $\Delta S_{max}(H)/\Delta S_{max}(0)$, with a sigmoidal function. This empirical fit captures the continuous crossover from the antiferromagnetic state at low field to the field-suppressed regime at high field, allowing a characteristic field scale, $H_c$, to be extracted. The extracted values increase systematically with thickness, from ≈ 40 mT in the bilayer to ≈ 85 mT in the 6-layer sample, indicating that the antiferromagnetic phase becomes progressively more robust as the effective interlayer exchange strengthens. These characteristic fields are substantially lower than the $\sim$200 mT required to suppress antiferromagnetic order in low-temperature magnetotransport measurements.[13,27] This difference is expected because the present calorimetric measurements are performed close to the Néel transition (120–135 K), where thermal fluctuations significantly weaken the interlayer antiferromagnetic coupling, reducing the magnetic field required to destabilize the ordered state. The smooth evolution of $\Delta S(H)$ shows that the suppression of antiferromagnetic order occurs through a continuous crossover rather than a sharp spin-flip transition within the explored field range. The applied field favors spin configurations along the easy axis, thereby suppressing magnetic fluctuations and the associated accumulated entropy change.

The magnetic entropy provides more than a measure of transition strength: by comparing the recovered entropy with the spin-only entropy of localized $Cr^{3+}$ moments, we can extract an entropy-derived effective magnetic moment per layer. This estimate should not be interpreted as a direct magnetometry measurement, but as the spin density whose full ($S$=3/2) entropy would correspond to the entropy recovered within the selected temperature window. Assuming localized $Cr^{3+}$ moments with ($S$=3/2), the spin-only entropy limit is $k_\mathrm{B}\ln(2S+1) = k_\mathrm{B}\ln(4)$ per Cr ion. The entropy-equivalent number of spins is therefore:

$$N_{eff} = \frac{\Delta S_{max}}{k_B \ln{(4)}} \tag{1}$$

Here, ΔS is obtained by integrating the excess heat capacity divided by temperature over the selected temperature window associated with the magnetic anomaly. Since each $Cr^{3+}$ ion carries a magnetic moment $gS\mu_B \approx 3\mu_B$, the corresponding effective magnetization can be written as:

$$M_{eff} = N_{eff} * 3\mu_B \tag{2}$$

Normalizing by the flake area $A$ and the number of layers $N_{layers}$, the entropy-equivalent areal moment density per layer is given by:

$$\sigma_{\mu,\text{layer}} = \frac{3\,\mu_B\,\Delta S_{\max}}{A\,N_{\text{layers}}\,k_B \ln(4)} \tag{3}$$

This quantity corresponds to an entropy-derived effective magnetic moment that captures the total magnetic degrees of freedom recovered within the experimental temperature range, rather than the full saturation magnetization.

The extracted values, shown in Figure 3f, increase with thickness and approach the spin-only saturation scale expected from the Cr areal density. Using the in-plane lattice parameters of CrSBr and two Cr atoms per in-plane unit cell gives $n_{\text{Cr}} \approx 2/(ab) \sim 12\,\text{nm}^{-2}$, corresponding to $\sigma_{\mu,\text{sat}} \approx n_{\text{Cr}} \cdot 3\mu_B \sim 36\,\mu_B/\text{nm}^2$ per layer. The approach toward this scale indicates that, in thicker flakes, a larger fraction of the available spin entropy contributes to the resolved thermodynamic response.[32] In thinner samples, the reduced entropy-derived effective moment is consistent with previous reports of thickness-dependent magnetization in few-layer CrSBr, where the moment per layer increases with layer number and approaches the bulk value only after several layers.[32,48] Overall, this entropy-based analysis provides a quantitative thermodynamic measure of how the magnetic degrees of freedom involved in interlayer ordering evolve with thickness.

## Conclusions

In conclusion, we demonstrate that microsecond pulse-heating nanocalorimetry provides direct, quantitative access to the magnetic thermodynamics of atomically thin van der Waals magnets. By measuring the heat capacity and magnetic entropy of individual CrSBr flakes under in-plane magnetic fields, we resolve how dimensionality, layer parity, and interlayer exchange govern magnetic ordering from the bulk-like regime down to the monolayer limit.

Our results reveal a systematic reduction of the $T^{*inter}$ with decreasing thickness, reflecting the weakening of interlayer antiferromagnetic coupling under dimensional confinement. They also uncover a pronounced parity effect: even-layer flakes exhibit a fully compensated antiferromagnetic ground state, whereas odd-layer flakes display an additional high-temperature thermodynamic signature associated with intralayer ferromagnetic correlations enhanced by an uncompensated net moment. In the monolayer, where interlayer antiferromagnetic order is absent, this ferromagnetic contribution remains essentially insensitive to moderate in-plane fields along the easy axis.

Beyond locating transition temperatures, the entropy analysis reveals how magnetic degrees of freedom are redistributed with thickness and magnetic field. A substantial fraction of the magnetic entropy is released over a broad temperature range extending above $T^{*inter}$, indicating that local magnetic correlations precede the establishment of long-range interlayer coherence. Applied magnetic field suppresses and redistributes this entropy, with a characteristic suppression field that increases from the bilayer to the 6-layer limit, consistent with stronger effective interlayer coupling in thicker flakes. By comparing the recovered entropy with the spin-only entropy expected for localized $Cr^{3+}$ moments, we further estimate an entropy-derived effective moment per layer, which increases with thickness and trends toward the expected spin-only saturation scale.

Together, these results establish a thermodynamic picture of dimensional crossover in CrSBr: reducing the layer number weakens interlayer antiferromagnetic coherence, enhances field susceptibility, and changes how intralayer magnetic correlations appear in the calorimetric response. More broadly, our work shows that nanocalorimetry can access heat-capacity anomalies, magnetic entropy and entropy-derived moment scales in individual exfoliated magnets, providing a thermodynamic complement to magnetotransport, optical and local magnetic probes. This approach should be applicable to other two-dimensional magnets and heterostructures in which strain, gating, stacking or engineered interlayer coupling are used to tune magnetic order.

## Methods

### Sample preparation

Bulk crystals of CrSBr were synthesized by chemical vapor transport (CVT) from a slightly off-stoichiometric mixture of chromium, sulfur, and chromium(III) bromide precursors sealed in an evacuated fused silica ampoule. The material was transported from the source zone at 950 °C to the growth zone at 850 °C, yielding high-quality layered crystals, following previously reported procedures[15].

CrSBr and hBN flakes were exfoliated from bulk single crystals onto $SiO_2$/Si substrates. Suitable CrSBr flakes were identified by optical contrast. The hBN/CrSBr heterostructures were assembled using a PVC-polymer dry-transfer technique[39] inside a $N_2$-filled glovebox and released onto the Pt heater/sensor of the calorimetric devices, avoiding any solvent-processing steps.[49] All exfoliation and transfer procedures were carried out in a nitrogen glovebox, $O_2$ < 0.5 ppm and $H_2O$ < 0.5 ppm, to prevent material degradation. The thickness of each CrSBr flake was measured by atomic force microscopy (AFM) as shown in the Supplementary Information.

**Device fabrication**

The calorimetric devices consisted of Ti/Pt strips (100 nm thick) patterned on 180 nm-thick SiN membranes using standard UV-Standard lithography and lift-off techniques. Each strip acted simultaneously as heater and thermometer, with a four-probe electrical configuration to minimize contact resistance. The suspended region of the membrane (15 × 15 $\mu m^2$ to 50 × 50 $\mu m^2$) ensured high thermal isolation and sub-picojoule sensitivity. Electrical leads and contact pads were fabricated from Ti/Pt (10/100 nm) and wire-bonded to the chip carrier for low-noise electrical connection. After fabrication, the devices were annealed at 250 °C in high vacuum to stabilize the Pt resistivity and remove adsorbates.

**Microsecond-Pulsed Heating Nanocalorimetry**

The heat capacity of CrSBr flakes was measured using microsecond-pulsed heating nanocalorimetry (µs-PHnC) in the temperature range from 80 to 300 K, allowing the phonon-dominated background to be constrained over a broader interval away from the magnetic anomaly. This improves the robustness of the lattice-background subtraction and reduces the uncertainty associated with the extraction of $\Delta C_p(T)$ and $\Delta S(T)$. The experiments were performed using custom-fabricated differential nanocalorimeters consisting of two suspended SiN membranes integrated in a series configuration: one acting as the reference and the other hosting the exfoliated flake. This differential approach enables direct subtraction of the addenda contribution (Pt and SiN membrane), allowing accurate extraction of the intrinsic heat capacity of ultrathin samples.

Each calorimetric cell incorporates a metallic Pt strip that serves simultaneously as heater and thermometer. Short current pulses (~50 µs) are applied to induce a controlled temperature increase, while the resulting resistance change is monitored in a four-probe configuration. The use of microsecond pulses minimizes lateral thermal diffusion, ensuring quasi-isothermal conditions within the sensing area and enabling reliable measurements on micrometer-scale flakes.

The heat capacity of the sample, $\Delta C_p(Ts(t))$, was extracted from the differential electrical response between the sample (S) and reference (R) using the following formula

$$\Delta C_p\big(T_S(t)\big) = \frac{I\Delta V}{\beta_S} - \frac{V_R}{\beta_S \beta_R}\frac{(d\Delta V/dt)_t}{(dR_S/dT_S)_t} + \frac{V_R I}{\beta_S}\left[1 - \frac{(dR_R/dT_R)_t}{(dR_S/dT_S)_t}\right] \quad (4)$$

The full derivation of the method is provided elsewhere.[37]

To isolate the magnetic contribution, the phonon background was subtracted from the measured heat capacity, yielding the excess heat capacity $\Delta C_p(T) = C_p(T) - C_{p,\mathrm{ph}}(T)$. The lattice contribution was modeled using a Debye expression with an adjustable scaling factor to account for deviations specific to ultrathin samples. The magnetic entropy change, ΔS, was then obtained by numerical integration of $\Delta C_p/T$ as a function of temperature. The lattice contribution is modeled according to the Debye expression for the heat capacity, including a proportionality factor A to account for deviations arising from the specific characteristics of the measured sample.

$$C_{p_p}^{ph}(T) = A\left(\frac{T}{T_D}\right)^3 \int_0^{T_D/T} \frac{t^4 e^t}{(e^t-1)^2} dt \qquad (5)$$

$T_D$ is the Debye temperature, and A is a scaling parameter determined from the best fit to the experimental data. Numerical integration of the Debye function enables accurate estimation of the lattice contribution, which is subsequently subtracted from the total $C_p(T)$ to isolate the excess heat capacity associated with the magnetic contribution.

The current density in the Pt heater is on the order of $10^5$ A/cm². This value is below typical thresholds required to induce spin–orbit torque effects in vdW magnetic systems.[50–53]

**Magnetic field configuration**

An in-plane magnetic field was applied during the calorimetric measurements using a custom-built copper coil. The coil generates magnetic fields up to 100 mT at its center, where the nanocalorimetric device is positioned. The field direction was aligned with the crystallographic $b$-axis of the sample, corresponding to the easy magnetization axis.

Since the magnetic response of CrSBr is mainly governed by the field component along the easy $b$-axis, the effective field used in the analysis was corrected by considering the angle between the applied field direction and the crystallographic $b$-axis of each flake. This angle was estimated from the optical orientation of the flake with respect to the device geometry. Unless otherwise stated, the magnetic-field values reported in the analysis correspond to the effective field projected along the $b$-axis.

The magnetic field was controlled through a precision current source, allowing a field resolution of ±2 mT. The field amplitude varied by adjusting the current supplied to the coil, ensuring stable and reproducible conditions during the measurements. A detailed characterization of the magnetic field as a function of the applied current and the position along the coil axis is provided in the Supplementary Information.

**Conflicts of Interest**

The authors declare no conflicts of interest.

**Acknowledgements**

The authors thank Prof. Jordi Sort for fruitful discussions and Arnau Villalobos-Martin for assistance with the AFM measurements. The ICN2 is funded by the CERCA program/Generalitat de Catalunya. The ICN2 is supported by the Severo Ochoa program of MINECO (Grant No. CEX2021-001214-S). H.G-T. acknowledges support from the Spanish Ministry of Science and Innovation through the FPI fellowship PRE2021-097883. This research has used the Spanish ICTS Network MICRO-NANOFABS, partially funded by FEDER funds through MINATEC-PLUS-2 project FICTS2019-02-40. The IMB-CNM-CSIC is supported by the María de Maeztu of MINECO (Grant No.CEX2023-001397-M). Synthesis work at Columbia University supported by the Air Force Office of Scientific Research (AFOSR) under MURI grant number FA9550-25-1-0288. P.G. acknowledges support from the F.R.S.-FNRS of Belgium (FNRS-CQ-1.C044.21-SMARD, FNRS-PDR-T.0128.24-ART-MULTI) and from the EU (ERC-StG-10104144-MOUNTAIN). K.W. and T.T. acknowledge support from the CREST (JPMJCR24A5), JST and World Premier International Research Center Initiative (WPI), MEXT, Japan. A. L and J.R.-V. acknowledge support from 2021SGR-00644 funded by Generalitat de Catalunya. This work was supported by the Spanish Ministry of Science, Innovation and Universities (MICIU) through projects PID2023-152783OB-I00, PID2023-149054NB- I00 and TED2021-129612B-C22.

**Data availability statement**

The dataset supporting the conclusions of this article has been deposited in CORA.RDR and is available through the following DOI: https://doi.org/10.34810/DATA3638.

**Table of contents**

Heat capacity and magnetic entropy of few-layer CrSBr are resolved for the first time, using pulse heating nanocalorimetry. Layer-dependent thermodynamic anomalies reveal a crossover from bulk-like antiferromagnetism to intralayer-dominated magnetic correlations, while magnetic fields suppress interlayer order. The entropy response provides a quantitative thermodynamic view of dimensionality, layer parity, and magnetic fluctuations in two-dimensional magnets.

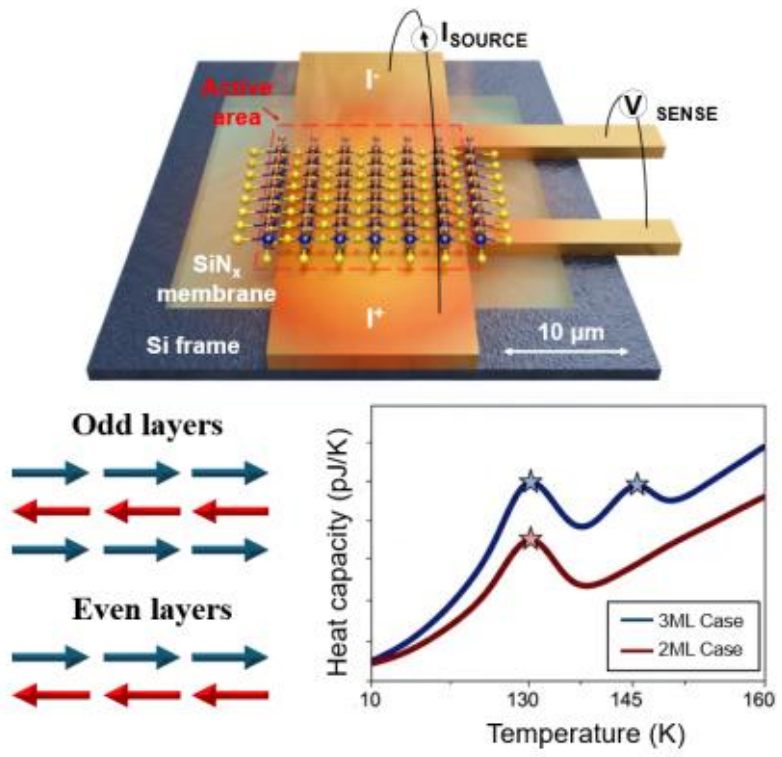

## Supplementary Information

# Layer- and Field-Dependent Magnetic Order in 2D CrSBr Revealed by Pulsed Nanocalorimetry

Hugo Gomez-Torres[1,2], Roop K. Mech[3], Alessandra Canetta[3], Llibertat Abad[4], Carles Navau[2], Daniel G. Chica[5], Xavier Roy[5], Pascal Gehring[3,6], Kenji Watanabe[7], Takashi Taniguchi[8], Aitor Lopeandia[1,2*], Javier Rodriguez-Viejo[1,2*]

[1] Catalan Institute of Nanoscience and Nanotechnology (ICN2), CSIC and BIST, E-08193, Bellaterra, Spain.
[2] Departament de Física, Universitat Autònoma de Barcelona, E-08193, Bellaterra, Spain.
[3] Institute of Condensed Matter and Nanosciences, Université Catholique de Louvain (UCLouvain), 1348 Louvain-la-Neuve, Belgium.
[4] Institut de Microelectrònica de Barcelona (IMB-CNM-CSIC). Campus de la UAB, E-08193, Cerdanyola del Vallès (Barcelona), Spain.
[5] Department of Chemistry, Columbia University, New York, New York 10027, United States.
[6] WEL Research Institute, Avenue Pasteur 6, 1300 Wavre, Belgium.
[7] Research Center for Electronic and Optical Materials, National Institute for Materials Science, 1-1 Namiki, Tsukuba 305-0044, Japan
[8] Research Center for Materials Nanoarchitectonics, National Institute for Materials Science, 1-1 Namiki, Tsukuba 305-0044, Japan

*Correspondence: Javier Rodriguez-Viejo (javier.rodriguez@uab.cat), Aitor Lopeandia (aitor.lopeandia@uab.cat)

## S1. Structural characterization of CrSBr flakes (AFM)

Atomic force microscopy (AFM) was used to determine the thickness and surface morphology of CrSBr flakes investigated in this work. Representative AFM topography images for bulk-like flakes, 6-layer (6ML) and 3-layer (3ML) samples are shown in Figure S1.

The thickness of each flake was extracted from height profiles taken across step edges. The measured step height between adjacent layers is ~0.8 nm, consistent with the interlayer spacing of CrSBr. The number of layers was determined by combining AFM measurements with optical contrast analysis.

For thicker flakes (bulk-like, 6ML and 3ML), the thickness was directly determined either from step-height profiles across the flake edges or, when step edges were not clearly accessible, from statistical analysis of the AFM height distribution. In the latter case, histograms of the topographic signal were constructed and fitted using Gaussian functions to resolve the height difference between regions corresponding to hBN and hBN/CrSBr, allowing an accurate extraction of the flake thickness.

For ultrathin flakes (2ML and monolayer), the thickness could not be reliably resolved by AFM due to the encapsulation within hBN and the limited vertical resolution in these conditions. In these cases, the number of layers was determined from optical contrast and deterministic transfer identification prior to encapsulation. This assignment is consistent with the expected thickness-dependent trends observed in the calorimetric measurements.

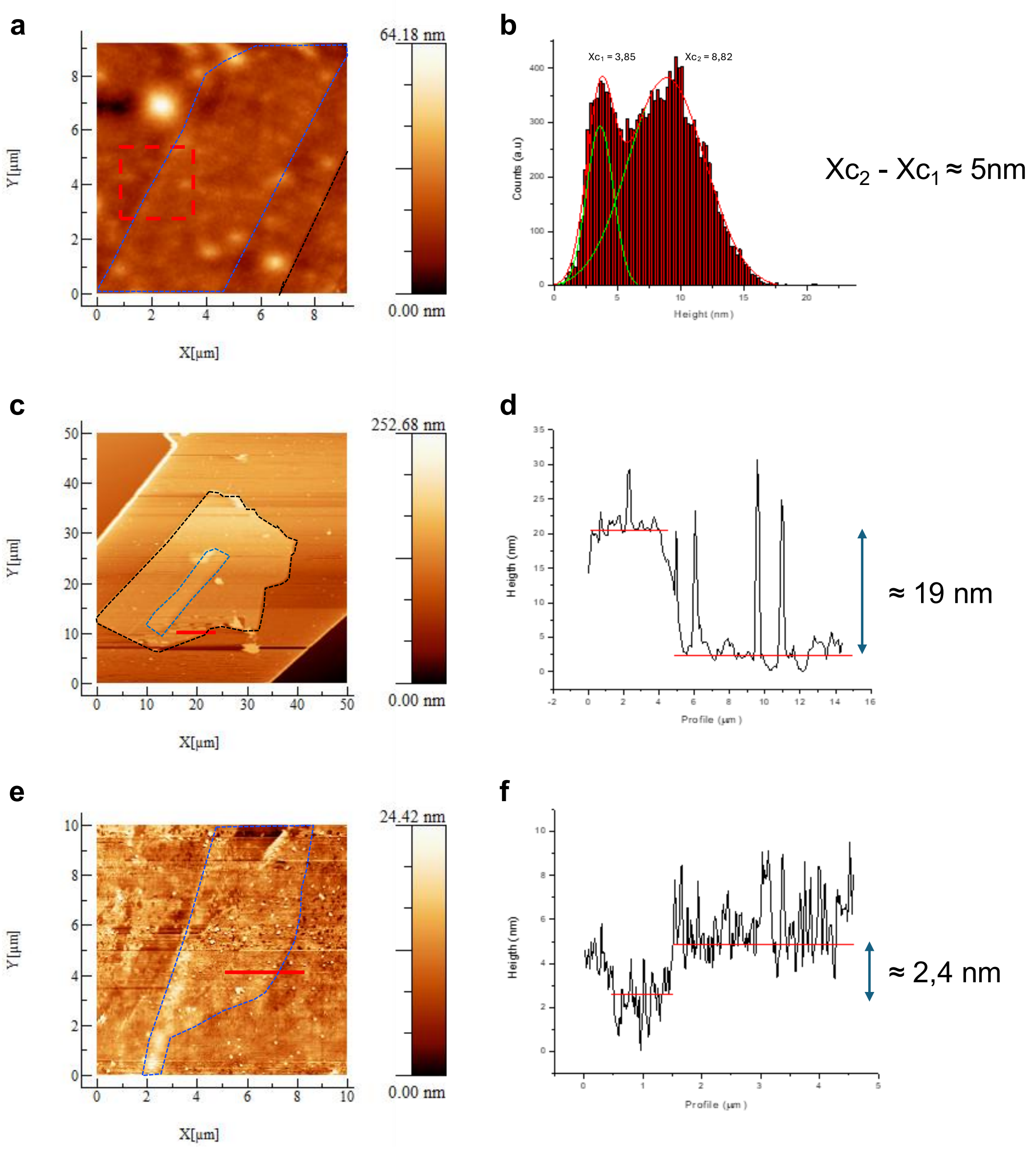

**Figure S1. AFM characterization of CrSBr flakes with different thicknesses.** (a,c,e) AFM topography images of CrSBr flakes: 6ML (a), bulk-like (c), and 3ML(e), encapsulated in hBN.

The outlines of the flakes are indicated by blue dashed lines, and representative regions used for thickness analysis are highlighted.

(b) Height histogram corresponding to the region marked in (a), showing two distinct contributions associated with hBN and hBN/CrSBr. Solid lines represent Gaussian fits used to extract the thickness difference. (d,f) Height profiles extracted along the lines indicated in (c) and (e), respectively. The step height between hBN and hBN/CrSBr regions provides the thickness of the flakes.

**S2. Determination of transition temperatures from Cp(T).**

The magnetic transition temperatures were extracted from the inflection points of the heat capacity curves $C_p(T)$, providing a model-independent and robust criterion for tracking the evolution of magnetic order across different thicknesses and magnetic fields.

To suppress high-frequency noise while preserving the intrinsic shape of the calorimetric anomaly, the experimental $C_p(T)$ data were fitted using a smoothing spline interpolation. Prior to fitting, the data were sorted in ascending temperature and inspected to remove possible outliers or non-physical points. The smoothing parameter was chosen to balance noise reduction and fidelity to the experimental signal. We verified that moderate variations of the smoothing parameter do not significantly affect the extracted transition temperatures (see below).

The transition temperature was determined from the inflection point of the $C_p(T)$ curve, defined as the temperature at which the second derivative vanishes:

$$\frac{d^2C_p}{dT^2} = 0$$

The derivatives were computed analytically from the spline fit. The zero crossing of $d^2C_p/dT^2$ was identified within a predefined temperature window centered around the calorimetric anomaly. This restriction avoids spurious inflection points arising from residual noise or secondary features in the data. To improve numerical accuracy, the exact position of the inflection point was obtained by linear interpolation between adjacent points where the sign of $d^2C_p/dT^2$ changes.

A simplified version of the Python routine used to extract the transition temperatures is shown below:

```
import numpy as np
from scipy.interpolate import UnivariateSpline
```

```
# Load data: T (K), Cp (J/K)
T, Cp = np.loadtxt("data.txt", unpack=True)
# Sort data
idx = np.argsort(T)
T, Cp = T[idx], Cp[idx]
# Spline smoothing
spline = UnivariateSpline(T, Cp, s=len(T)*np.var(Cp)*1e-4)
# Fine temperature grid
T_fine = np.linspace(T.min(), T.max(), 2000)
# Second derivative
d2 = spline.derivative(2)(T_fine)
# Find zero crossing (inflection point)
idx_zero = np.where(np.diff(np.sign(d2)))[0]
T_inflection = T_fine[idx_zero]
```

In the full implementation, the search for the inflection point is restricted to a temperature window around the magnetic transition, and the final transition temperature is selected as the zero crossing closest to the extremum of $dC_p/dT$.

The uncertainty in the extracted transition temperatures arises from experimental noise, numerical differentiation, and the choice of smoothing parameters. We estimate an uncertainty of ±(0.5–1) K. Importantly, the observed trends in $T^{*inter}$ as a function of thickness and magnetic field are significantly larger than this uncertainty, ensuring the robustness of the conclusions. For odd-layer samples, when the interlayer antiferromagnetic anomaly is suppressed at high field, the remaining high-temperature inflection point is assigned to the intralayer ferromagnetic contribution.

| Layer | Field (mT) | Transition | Extracted transition temperature (K) |
|---|---|---|---|
| 1 | 0<br>100 | Intralayer | 145 |
| 2 | 0 | Interlayer | 133,28 |
| | 28 | | 132,9 |
| | 33 | | 132,3 |
| | 40 | | 131,7 |
| | 100 | - | - |
| 3 | 0 | Interlayer | 134,3 |
| | 40 | | 133,3 |
| | 50 | | 131,4 |
| | 60 | | 130,2 |
| | 70 | | 129,4 |
| | 100 | Intralayer | 146 |
| 6 | 0 | Interlayer | 139,22 |
| | 55 | | 137,78 |
| | 61 | | 136,68 |
| | 66 | | 135,9 |
| | 69 | | 135,22 |
| | 72 | | 134,7 |
| | 75 | | 134,2 |
| | 80 | | 132,2 |
| | 100 | - | - |
| Bulk-like | 0 | Interlayer | 140,62 |
| | 100 | | 140,34 |

**Table S1. Extracted magnetic transition temperatures as a function of layer number and magnetic field.** Characteristic temperatures obtained from the inflection-point analysis of $C_p(T)$ for CrSBr flakes under different in-plane magnetic fields. The assigned magnetic feature indicates whether the anomaly is attributed to intralayer ferromagnetic-like correlations or to the interlayer antiferromagnetic transition.

**S3. Addenda and hBN background correction**

The measured heat capacity contains contributions from the CrSBr flake, the nanocalorimetric platform, and the encapsulating hBN layers. The addenda contribution of the calorimetric platform was first removed using the reference calorimeter measured under the same experimental conditions. This differential configuration minimizes common contributions from the SiN membrane, Pt heater, electrical wiring and slow temperature drifts.

After addenda subtraction, a residual smooth background remains due to the hBN encapsulation. This contribution varies from device to device because the hBN flakes used for encapsulation have different thicknesses and lateral areas. As a result, the absolute heat-capacity level can differ between samples even when the CrSBr contribution is normalized by flake area and number of layers.

To make the heat-capacity curves directly comparable, we corrected this residual background by applying a constant offset to each dataset. The offset was determined by matching the heat-capacity level in a reference temperature region well above the magnetic anomaly, where the hBN contribution is smooth and the magnetic contribution varies only slowly. The estimated hBN contribution and the corresponding offset applied to each dataset are summarized in Table S2.

The corrected heat capacity is therefore expressed as:

$$C_p^{CrSBr}(T) = C_p^{meas}(T) - C_p^{addenda}(T) - C_p^{hBN,offset}$$

where $C_p^{hBN,offset}$ accounts for device-to-device variations in the smooth hBN background arising from differences in encapsulation thickness and area.

Importantly, this correction only shifts the absolute heat-capacity baseline and does not modify the shape, temperature position or field dependence of the magnetic anomalies. All field-dependent comparisons are performed within the same device, so they are unaffected by the device-to-device hBN offset correction. We verified that reasonable variations of the applied offset change the absolute magnitude of $C_p$ and of the extracted magnetic entropy, but do not affect the determination of $T^{*inter}$ or the observed trends with thickness and magnetic field. Figure S2 illustrates the hBN-related baseline correction applied to compare different CrSBr devices. The heat-capacity contribution of the hBN encapsulation was estimated from the measured hBN thickness and lateral area of each device, using reported heat-capacity values for hBN in the relevant temperature range [Phys. Rev. B **13**, 4607]. Because the hBN contribution is smooth and non-magnetic over the temperature window studied here, it was treated as a constant offset for comparison purposes. This offset was subtracted from each dataset to compensate for device-to-device differences in hBN encapsulation volume. The correction aligns the absolute heat-capacity baselines of the different samples, while preserving the shape, position, and field dependence of the magnetic anomalies.

| Sample | CrSBr area | hBN thickness/area | Estimated Cp hBN contribution | Applied offset |
|---|---|---|---|---|
| 1ML | 37,5 µm$^2$ | 21 nm / 123 µm$^2$ | 1.63 pJ/K | - 1,63 pJ/K |
| 2ML | 47,5 µm$^2$ | 25 nm / 310 µm$^2$ | 4.88 pJ/K | - 4,88 pJ/K |
| 3ML | 35 µm$^2$ | 34 nm / 650 µm$^2$ | 13,9 pJ/K | - 13,9 pJ/K |
| 6ML | 67 µm$^2$ | 11 nm / 290 µm$^2$ | 2,01 pJ/K | - 2,01 pJ/K |
| Bulk (25ML) | 90 µm$^2$ | 39 nm / 870 µm$^2$ | 21,4 pJ/K | - 21,4 pJ/K |

**Table S2. hBN-related heat-capacity offset correction.** Summary of the geometrical parameters of the hBN encapsulation and the constant heat-capacity offset applied to each CrSBr device. The correction accounts for differences in hBN thickness and lateral area between devices. Since hBN provides a smooth, non-magnetic heat-capacity background in the investigated temperature range, the offset affects the absolute Cp level but not the position or field dependence of the magnetic anomalies.

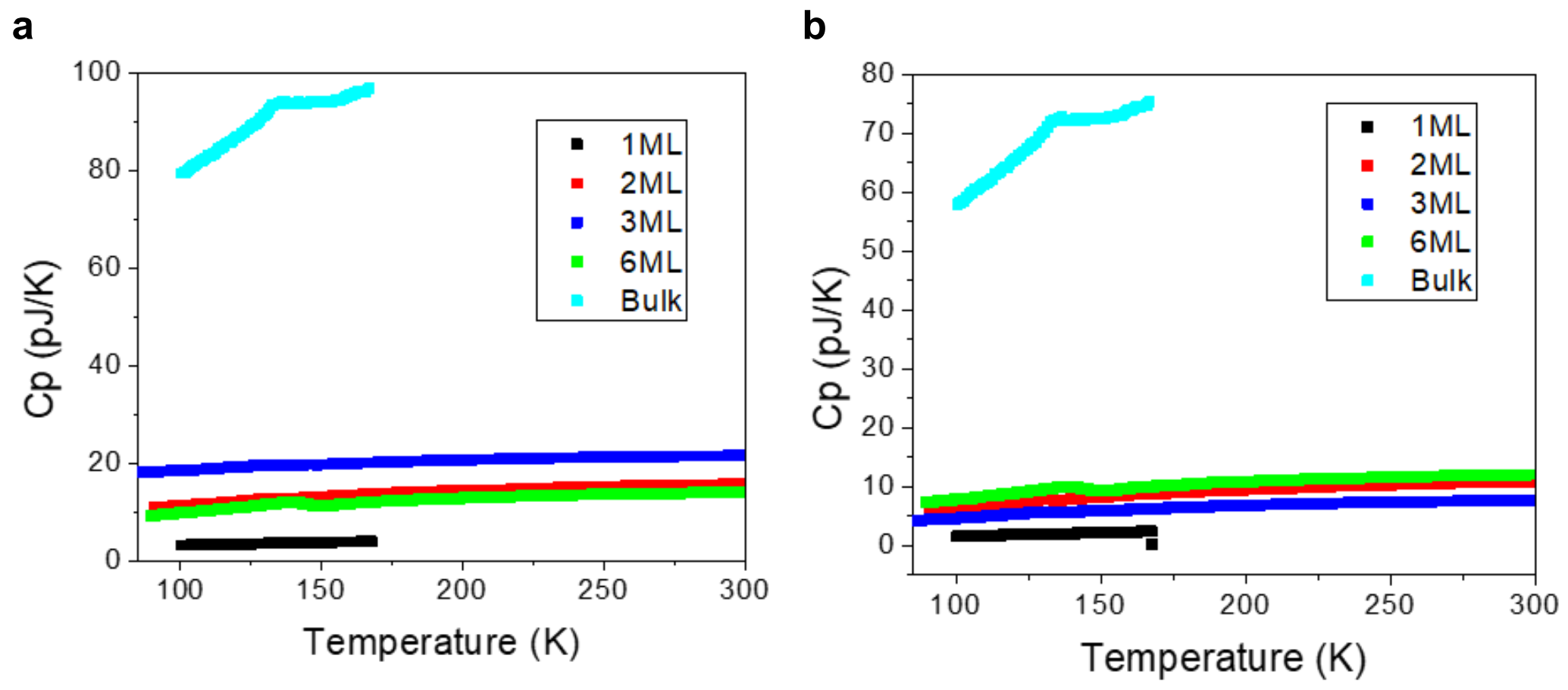

**Figure S2. Heat capacity correction for different CrSBr devices.** (a) Measured heat-capacity contribution for each encapsulated CrSBr Device. (b) Corrected heat-capacity curves after subtraction of the hBN-related constant offset. The correction shifts the absolute baseline but does not affect the position or field dependence of the magnetic anomalies.

**S4. Phonon background subtraction**

To isolate the magnetic contribution to the heat capacity, the lattice background was subtracted from the measured $C_p(T)$curves. This step is necessary because the total heat capacity is dominated by phonons over the full temperature range, while the magnetic anomaly associated with the onset of magnetic order appears as a smaller superimposed contribution.

For each CrSBr flake, the phonon contribution was modeled using a Debye-type expression,

$$C_{\mathrm{ph}}(T) = A \cdot 9Nk_B\left(\frac{T}{\Theta_D}\right)^3 \int_0^{\Theta_D/T} \frac{x^4 e^x}{(e^x-1)^2}\,dx$$

where $N$is the number of oscillators, $k_B$is the Boltzmann constant, $\Theta_D$is the effective Debye temperature, and $A$ is a scaling factor introduced to account for sample-dependent effects such as thickness, finite-size deviations, and uncertainties in the absolute background level. In practice, $A$ and $\Theta_D$ were treated as effective fitting parameters. Therefore, the Debye-type function is used to reproduce the smooth phonon-dominated background rather than to provide a complete microscopic description of the phonon density of states.

The fitting was performed in a temperature region above the main magnetic anomaly, where the heat capacity is dominated by the lattice contribution and the magnetic signal becomes smooth and weak. The resulting Debye background was then extrapolated over the full temperature range and subtracted from the experimental $C_p(T)$ data to obtain the excess heat capacity. For a given device, the same phonon background was used for all magnetic-field values, since the lattice contribution is field independent within the explored field range.

The excess heat capacity was then calculated as:

$$\Delta C_p(T) = C_p(T) - C_{\text{ph}}(T)$$

which is assigned to the magnetic contribution.

The magnetic entropy was subsequently obtained from:

$$\Delta S(T) = \int_{T_0}^{T} \frac{\Delta C_p(T)}{T} \, dT$$

where $T_0$ was chosen below the onset of the magnetic anomaly. The upper integration limit extends well above the transition temperature in order to capture the broad entropy release associated with short-range magnetic correlations.

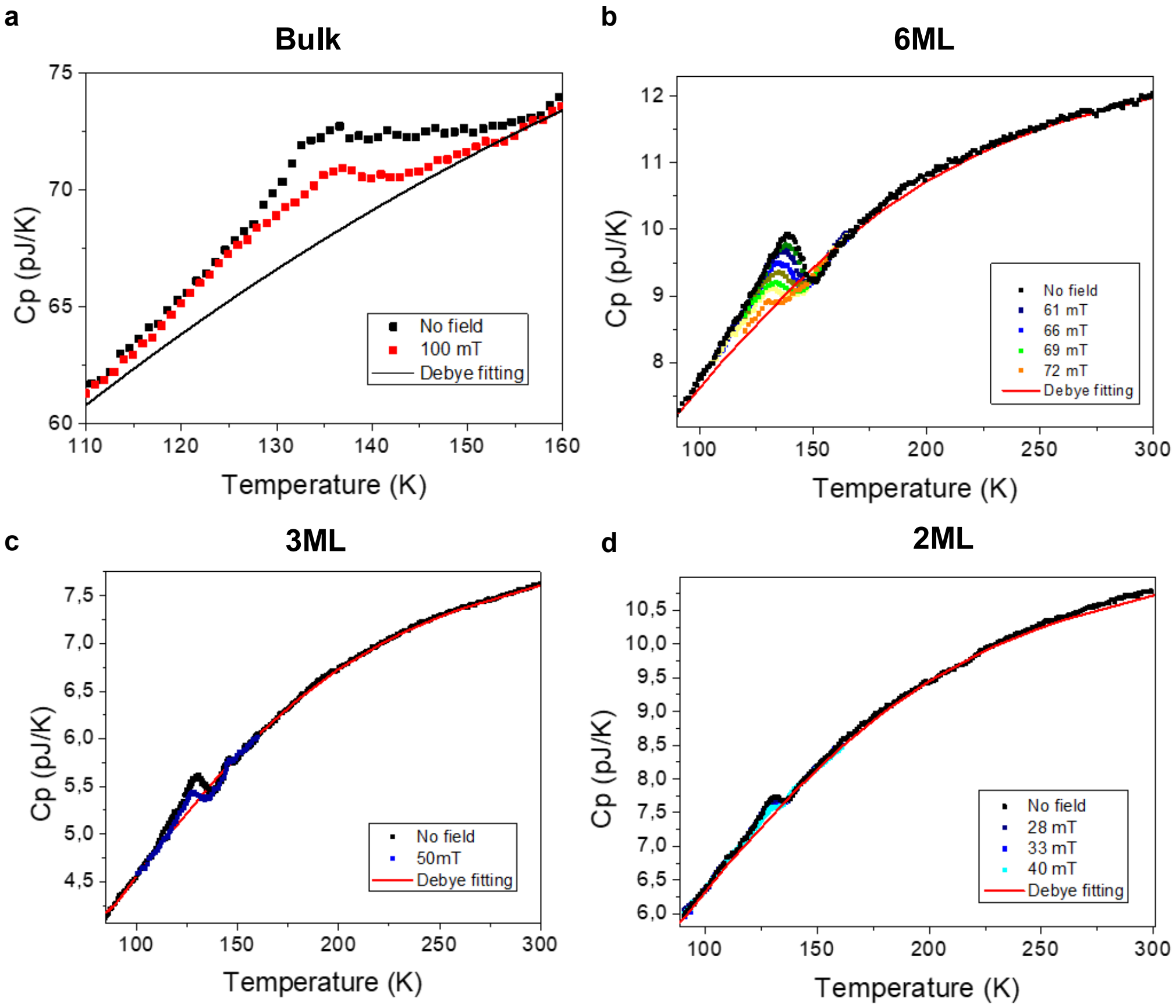

**Figure S3. Effective phonon background contribution for CrSBr flakes.** (a-d) Measured heat capacity $C_p(T)$ for bulk-like, 6ML, 3ML and 2ML, together with the corresponding Debye fits $C_{\mathrm{ph}}(T)$.

Figure S3 shows the experimental $C_p(T)$ curves for the samples exhibiting a resolvable interlayer antiferromagnetic anomaly, together with the corresponding effective Debye-type backgrounds. The subtraction reveals a clear thickness-dependent evolution of the magnetic contribution: the excess heat capacity decreases in magnitude, broadens, and shifts to lower temperature as the number of layers is reduced.

To assess the robustness of the analysis, the fitting range and Debye parameters were varied within reasonable bounds. The fitting was performed over the extended 80–300 K range, with the magnetic-anomaly region excluded from the fit. The high-temperature part of the curve provides the main constraint on the smooth lattice background, while the region around the magnetic transition was excluded to avoid subtracting magnetic entropy as part of the phonon contribution.

## S5. Magnetic field calibration of the coil

The magnetic field applied during the calorimetric measurements was generated using a custom-built copper coil integrated in the cryogenic setup. Accurate calibration of the field amplitude and spatial distribution is essential to ensure reliable determination of the field dependence of the heat capacity and magnetic entropy.

The magnetic field $H$ at the position of the sample was calibrated as a function of the current $I$ supplied to the coil using a calibrated Hall probe placed at the center of the coil. The measurements were performed under the same geometrical conditions as the calorimetric experiments.

A linear dependence between magnetic field and applied current was observed over the full range of operation:

$$H = \alpha I$$

where $\alpha$ is the coil calibration factor. From the calibration, we obtain: 12.1 mT/A. This linear behavior, as shown in Figure S4, confirms the absence of significant nonlinearities in the field-current response within the explored current range.

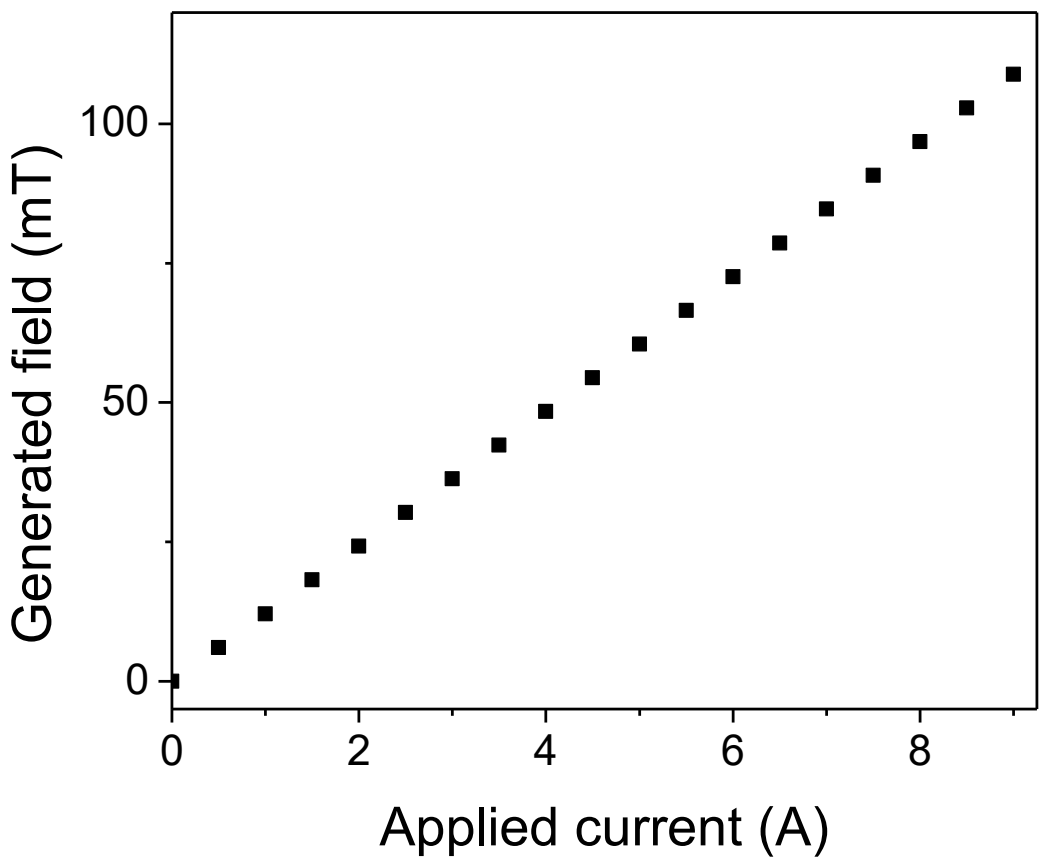

**Figure S4.** Generated magnetic field as a function of the applied current, measured using a calibrated Hall probe at the sample position.

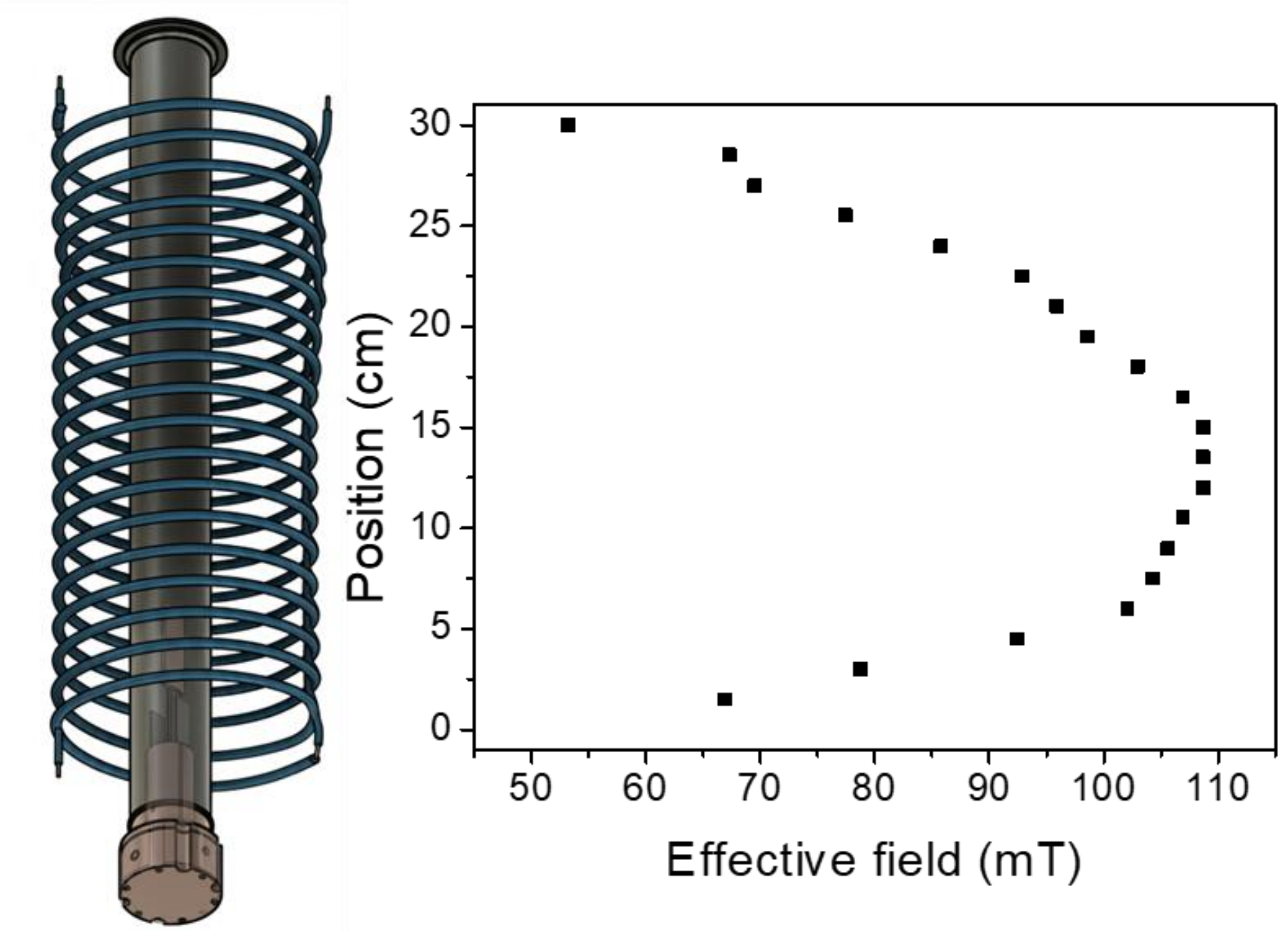

**Figure S5. Magnetic field calibration and spatial profile of the coil at 9 A.** Schematic of the custom-built copper coil used to generate the in-plane magnetic field during calorimetric measurements, together with the measured magnetic-field profile along the coil axis. The maximum field is obtained near the central region of the coil, where the nanocalorimetric device is positioned during the experiment.

Figure S5 shows the coil geometry and the magnetic-field profile along the coil axis. The nanocalorimetric device was positioned close to the maximum-field region. Although the field varies over the centimeter-scale length of the coil, its variation over the micrometer-scale dimensions of the CrSBr flakes and calorimetric cell is negligible. The magnetic field was applied in-plane and aligned with the crystallographic $b$-axis of CrSBr, corresponding to the magnetic easy axis.